\documentclass[sigconf]{acmart}
\usepackage{amssymb}
\usepackage{color}
\usepackage{balance}
\usepackage{todonotes}
\usepackage{listings}
\usepackage{amsmath}
\usepackage{booktabs}
\usepackage{algorithm}
\usepackage{algorithmicx}
\usepackage{algpseudocode}
\usepackage{multirow}
\usepackage{soul}
\usepackage{url}
\usepackage{xfrac}
\usepackage{graphicx}
\usepackage{subfig}
\usepackage{tabularx}
\usepackage{amsmath}
\usepackage{pifont}
\usepackage{enumitem}
\usepackage{setspace}

\usepackage{adjustbox}
\usepackage{diagbox}
\usepackage{tabularx,ragged2e,booktabs}
\usepackage{arydshln}

\setcopyright{acmcopyright}

\makeatletter
\def\@copyrightspace{\relax}
\makeatother
\begin{document}
\title{Detecting, Understanding and Supporting \\* Everyday Learning in Web Search}

\author{Ran Yu}
\affiliation{%
  \institution{L3S Research Center}
  \streetaddress{Appelstr. 4}
  \city{Hannover} 
  \country{Germany} 
  \postcode{30167}
}
\email{yu@l3s.de}

\author{Ujwal Gadiraju}
\affiliation{%
  \institution{L3S Research Center}
  \streetaddress{Appelstr. 4}
  \city{Hannover} 
  \country{Germany} 
  \postcode{30167}
}
\email{gadiraju@l3s.de}

\author{Stefan Dietze}
\affiliation{%
  \institution{L3S Research Center}
  \streetaddress{Appelstr. 4}
  \city{Hannover} 
  \country{Germany} 
  \postcode{30167}
}
\email{dietze@l3s.de}


\begin{abstract}
Web search is among the most ubiquitous online activities, commonly used to acquire new knowledge and to satisfy learning-related objectives through informational search sessions. The importance of learning as an outcome of web search has been recognized widely, leading to a variety of research at the intersection of information retrieval, human computer interaction and learning-oriented sciences. Given the lack of explicit information, understanding of users and their learning needs has to be derived from their search behavior and resource interactions. In this paper, we introduce the involved research challenges and survey related work on the detection of learning needs, understanding of users, e.g. with respect to their knowledge state, learning tasks and learning progress throughout a search session as well as the actual consideration of learning needs throughout the retrieval and ranking process. In addition, we summarise our own research contributing to the aforementioned tasks and describe our research agenda in this context. 
\end{abstract}

%
%



\maketitle

\section{Introduction}
Web search is among the most frequent online activities and has become a ubiquitous task. As is common search practice, a coherent search \textit{session}, involving a particular search intent, usually involves several queries as well as one or more breaks in between (cf. ~\cite{hagen2013}).

In particular, \textit{informational} search sessions~\cite{broder2002}, i.e. sessions pertaining to the search for a particular piece of information expected to be available on the Web, are common and involve a particular learning intent, that is, the intent to acquire knowledge with respect to a certain topic. 

Whereas platforms dedicated to online learning, such as MOOC environments, are tailored towards improving the learning performance and experience of online users, contemporary search engines have to satisfy a range of use cases, which may or may not involve learning. \textit{Transactional} search sessions~\cite{broder2002} are a common example of non-learning related online search. In contrast to actual learning-oriented environments in the online or offline sphere, where certain knowledge about the learning intent, the user as well as the learning task usually is available, such information is lacking in general online search settings. Consequently, heterogeneous features observable throughout a Web search session have to be utilised to derive insights about the learning intent, the user and the actual learning task. 

Recently, a range of research works have approached this problem, often summarised unter the `\textit{search as learning (SAL)}' umbrella and involving distinct disciplines such as information retrieval, human computer interaction or machine learning. 

This paper attempts to provide an overview of a SAL research agenda by (i) summarising research challenges involved in this context, (ii) discussing related works in the area, (iii) presenting insights into early results of the authors' own work as well as (iv) introducing gaps and future work in this area.


\begin{figure*}[!htbp]
\centering
\includegraphics[clip=true, trim=2pt 0pt 4pt 4pt, width=0.9\textwidth]{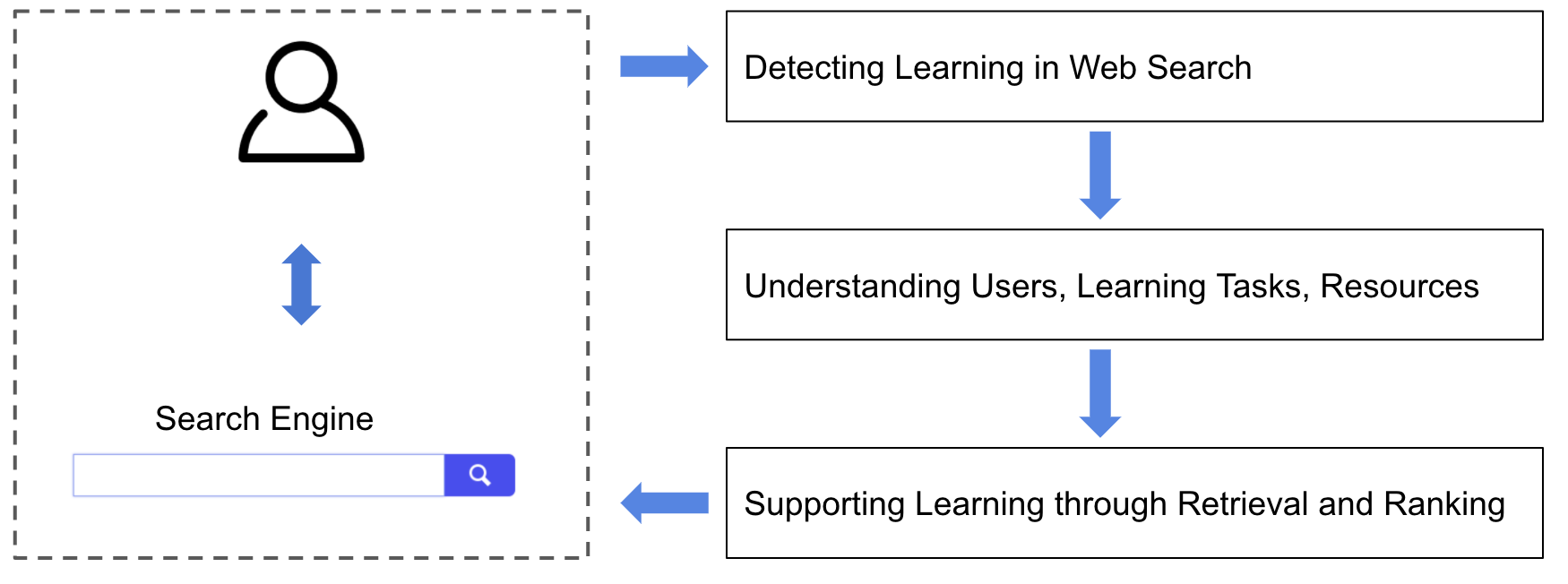}
\caption{The detecting, understanding, supporting everday learning in Web search pipeline.}
\label{fig:pipeline}
\end{figure*}

Figure ~\ref{fig:pipeline} summarises the key emerging research challenges which at the same time define the structure for the remaining sections. \textit{Detecting Learning in Web Search} (Section \ref{sec:detect}), refers to the process of distinguishing learning-related activities from other, non-learning, activities in general Web search scenarios. \textit{Understanding Users, Learning Tasks, Resources} (Section \ref{sec:understand}) refers to the challenges involved in inferring information about a user, such as his/her knowledge state, the learning task, such as its complexity, or the involved resources from unstructured behavioral data observable throughout an online search session. Finally, \textit{Supporting Learning through Retrieval and Ranking} (Section \ref{sec:support}) refers to the actual consideration of inferred learning needs as part of the retrieval and ranking process or through adapting search interfaces to the user's learning intent.


\section{Detecting Learning in Web Search} \label{sec:detect}


Whereas only a certain amount of Web search sessions include a particular learning need, identifying such sessions becomes a prerequisite to facilitate further applications for understanding and supporting learning. 

An established taxonomy from Broder~\cite{broder2002} that has been widely used in the Web search context distinguishes between transactional, navigational and informational search sessions, where in particular the latter involve a learning goal, i.e. the intent to acquire knowledge about a particular topic. 
Specifically, \textit{transactional} search sessions usually aim at conducting a specific online transaction, such as, purchasing a ticket, \textit{navigational} queries merely are aimed at leading the user to a dedicated website. In contrast, \textit{informational} sessions imply the intent of a user to acquire some information assumed to be present on one or more web pages. In this context, the same query, for instance, \textit{Elbphilharmonie} may be used to either buy tickets in a particular concert venue (transactional), to reach the Website \url{https://www.elbphilharmonie.de} (\textit{navigational}) or to acquire knowledge about the \textit{Elbphilharmonie} (informational). We inspected a real-world query log, which consists of 913 search sessions, we found that 49.7\% of them were informational search sessions with specific learning intent.

By adopting Broder\'s taxonomy, the task of detecting learning in Web search can be formalised as identifying informational Web search sessions. 
Here a Web search \textit{session} refers to the search activities within a certain time period that share the same information need. Previous work~\cite{hagen2013} on segmenting such search sessions have achieved promising performance. Here we do not go into details about the session segmentation task but focus on the automated detection of the intent of Web search.

\subsection{Related Works}

The classification of Web search queries has been explored in several different scenarios, for instance, to classify a query into one of the categories~\cite{kang2003,lee2005,liu2006,kathuria2010} or subcategories~\cite{jansen2008} of Broder\'s taxonomy, or into other classes that are tailored towards specific applications~\cite{yates2006,hu2009,kravi2016one}.

Early studies on intent classification relied on manual approaches, for instance, by asking users through surveys~\cite{broder2002} or by manual annotation of intents through judges~\cite{rose2004}. However, while this process does not scale well to large datasets, automated classification approaches have been explored. Both supervised~\cite{kang2003,lee2005,liu2006,yates2006,jansen2008,hu2009,kravi2016one} and unsupervised~\cite{yates2006,kathuria2010} approaches have been applied on the classification of Web search queries. 
The features utilised in the aforementioned approaches are extracted from query terms~\cite{jansen2008,hu2009,kathuria2010,kravi2016one}, user-click behaviors~\cite{lee2005,liu2006,kravi2016one}, anchor-links~\cite{lee2005,liu2006}, Web documents' content~\cite{jansen2008} and page views~\cite{kathuria2010,kravi2016one}. 

The aforementioned works focus on the classification of single query sessions, often limited to data collected through lab studies. However, recent studies have shown that users information seeking tasks have grown more sophisticated~\cite{jones2008beyond} and often require one or more queries across multiple search sessions~\cite{kotov2011,liu2010,agichtein2012}.

\subsection{Detecting the Intent of Web Search Sessions from User Interaction Features}\label{sec:detect_approach}

In contrast to such previous works, focused on query-based intent detection, our ongoing work to address this problem focuses on automatically detecting the intent of search activities at the level of search sessions. 
 
\textbf{Approach. }
We approach the problem of detecting informational Web search sessions with supervised models for classification. We extract 22 features according to multiple dimensions of a search session, structured into three categories, namely features related to \textit{Query} (i.e.features related to number of query terms and the between query similarity), \textit{Session} (i.e. total number of queries issued, session duration related and session breaks related features) and \textit{Browsing} behaviour (i.e. features related to number of clicks, revisited pages and similarity between query and the clicked URL). 
For the classification model, we have experimented with several different approaches. Considering the scale of the data as well as the number and characteristics of the features, we have opted for Decision Tree (DT), Logistic Regression (LR), Support Vector Machine (SVM)~\cite{platt1999} and Random Forest (RF)~\cite{breiman2001} as classification models. We tune the hyper parameters of each classifier through grid search. The preliminary result of the performance of each classifier is reported below.

\textbf{Preliminary Result. }

\begin{table*}[h]
	\caption{Performance of different classifiers.}
	\label{tab:eva_classifier}
	\small
	\centering
 	\scalebox{1}{
	\begin{tabular}{p{1cm}| p{0.5cm}p{0.5cm}p{0.5cm}|p{0.5cm}p{0.5cm}p{0.5cm}|p{0.5cm}p{0.5cm}p{0.5cm}|p{0.5cm}p{0.5cm}p{0.5cm}|p{0.5cm}p{0.5cm}p{0.5cm}|p{0.5cm}}
	\toprule
   & \multicolumn{3}{c}{\textbf{Navigational}}& \multicolumn{3}{c}{\textbf{Informational}}& \multicolumn{3}{c}{\textbf{Transactional}}& \multicolumn{3}{c}{\textbf{Weighted average}} & \textbf{All} \\
   \textbf{Method} & \textbf{P}& \textbf{R}& \textbf{F1}& \textbf{P}& \textbf{R}& \textbf{F1}& \textbf{P}& \textbf{R}& \textbf{F1}& \textbf{P}& \textbf{R}& \textbf{F1} &\textbf{Accu} \\
   \toprule
   
 DT & 0.764 & 0.731 & 0.747 & 0.644 & 0.839 & 0.728 & 0.241 & \textbf{0.076} & \textbf{0.116} & 0.599 & 0.653 & 0.611 & 0.653\\
 SVM & 0.786 & \textbf{0.760} & \textbf{0.773} & \textbf{0.656} & 0.927 & 0.768 & \textbf{0.800} & 0.022 & 0.042 & \textbf{0.724} & \textbf{0.694} & 0.623 & \textbf{0.694}\\
LR & \textbf{0.809} & 0.709 & 0.756 & 0.651 & \textbf{0.938} & \textbf{0.769} & 0.556 & 0.054 & 0.099 & 0.680 & 0.691 & \textbf{0.630} & 0.691\\
RF & 0.782 & 0.731 & 0.756 & 0.648 & 0.923 & 0.761 & 0.556 & 0.027 & 0.052 & 0.670 & 0.685 & 0.617 & 0.685\\

   \bottomrule
	\end{tabular}}
\end{table*}
We apply our model to a dataset of real-world query logs, which contains 6860 queries from 124 users corresponding to 913 sessions. Each session has been manually annotated by at least two annotators and assigned to one of the three classes. The annotated dataset is available online\footnote{http://l3s.de/\textasciitilde yu/mission\_classification}.
The results of using standard precision, recall and F1 score for each individual class, as well as their average across classes and the overall accuracy of the tested classifiers are shown in Table \ref{tab:eva_classifier}. For all configurations, the classification accuracy is above 0.653 and the F1 score is above 0.611, which indicates that the set of features we extracted from user search activities can provide meaningful evidence for detecting the search intent.

We also analyzed the information gain of the selected features, and found that features in the browsing category appear more important than features in other categories, with 2 browsing features ranking at the top 2 positions. Query features are also shown to be effective with 3 features among the top 6. Session-based features have the least contribution among all 3 categories. 

For simplicity, we use the term ``session'' in the remaining of this paper to refer to informational Web search sessions in particular, i.e. sessions which involve a particular learning intent.

\subsection{Future Work}
The overall classification performance indicates reasonable results on average, in particular transactional sessions appear ambiguous for both human annotators as well as supervised models. For this reason, results indicate that more specific classification tasks are likely to yield superior performance. For instance, an application-specific classifier aimed at targeted advertising may focus on only transactional or informational sessions (depending on the advertised offering), so that binary classification can be applied through a more tailored model. 
Further more, we found limitations arise in particular from the nature of the experimental dataset and the lack of publicly available, up-to-date query logs, future work will be concerned in particular with the application of similar approaches on a more recent and larger scale dataset. This would enable supervised models which are better reflecting contemporary search behavior and at the same time, utilise a wider variety of features. 
\section{Understanding the Learning Process and Outcomes} \label{sec:understand}

Since a sound understanding of learning throughout the search process is required in order to support learning, in this section we summarize the existing efforts in relevant topics and introduce our ongoing works on this task. 

There are several factors that potentially affect the learning performance and the required support throughout the search session. These can be classified into three main classes, namely, \textit{user}, e.g. the initial knowledge state and behavioural pattern, \textit{learning task}, such as the task difficulty and novelty, and \textit{resource}, e.g. the complexity or relevance of a resource. These factors are strongly inter-dependent. For instance, the task difficulty is subjective to user's knowledge state, whereas the user's knowledge state is a decisive factor on the resource selection.

Given the sparsity and heterogeneity of data throughout a search session, the works discussed in this section aim at inferring the aforementioned notions by considering a range of features observable throughout a search session. 


\subsection{Related Works}
Previous works assessed the relation between learning and user search behavior from several different perspectives. 
Eickhoff et al.~\cite{eickhoff2014lessons} investigated the correlation between features extracted from search session as well as search engine result page (SERP) documents with learning needs related to either procedural or declarative knowledge. 
The influence of distinct query types on knowledge gain was studied by Collins-Thompson et al.~\cite{collins2016assessing}, finding that intrinsically diverse queries lead to increased knowledge gain. 
%
Hagen et al.~\cite{hagen2016writers} investigated the relation between the writing behavior and the exploratory search pattern of writers and revealed that query terms can be learned while searching and reading. 
Vakkari~\cite{vakkari2016searching} provided a structured survey of features indicating learning needs as well as user knowledge and knowledge gain throughout the search process. 
Zhuang et al.~\cite{zhuang2017understanding} investigated the possibility of using 37 user search behavioral features to predict the user engagement, which correlates with learning, with supervised classifiers. 
By matching the learning tasks into different learning stages of Anderson and Krathwohl's taxonomy~\cite{anderson2001taxonomy}, Jansen et al. studied the correlation between search behaviors of 72 participants and their learning stage \cite{jansen2009using}. 
Gwizdka et al.~\cite{gwizdka2016towards} proposed to assess learning outcomes in search environments by correlating individual search behaviors with corresponding eye-tracking measures. 
%
White et al.~\cite{white2009characterizing} investigated the difference between the behavior of domain experts and non-experts in seeking information on the same topic. By analyzing the activity log of experts and non-experts across different domains, the authors found that the distribution of features such as number of queries and query length differed across the levels of expertise. 
Zhang et al. ~\cite{zhang2015predicting, zhang2011predicting} explored using search behavior as an indicator for the domain knowledge of a user based on data acquired through a lab study ($n=35$). 
Further, Cole et al.~\cite{cole2013inferring}, observed that behavioral patterns provide reliable indicators about the domain knowledge of a user, even if the actual content or topics of queries and documents are disregarded entirely.
Gwizdka and Spence~\cite{gwizdka2006can} have shown that a searcher's perception of task difficulty is a subjective factor that depends on the domain knowledge and some other individual traits. 
Arguello~\cite{arguello2014predicting} proposed to use logistic regression to predict task difficulty in a search environment using behavioural features. 

The aforementioned prior works have either studied a limited set of features or have addressed only specific learning scenarios and learning types. In particular, the generalizability of knowledge gain measures in previous works has not been investigated. 

\subsection{Analyzing Knowledge Gain of Users in Informational Search Sessions on the Web} \label{sec:analyse}
We extend the current understanding of user knowledge gain in informational search sessions. Using real world information needs and search sessions on the Web, we investigate the possibility of using search activity related features to predict knowledge gain (Section \ref{sec:predict}).

In particular, our recent work~\cite{gadiraju2018chiir} investigated the impact of information needs on the search behavior and knowledge gain of users. To further the current understanding of the impact of informational search on a user's knowledge, we recruited 500 distinct users from a crowdsourcing platform and orchestrated search sessions spanning 10 different information needs. We followed the recommended guidelines for effective crowdsourcing \cite{gadiraju2015understanding,gadirajuclarity}. By employing scientifically formulated knowledge tests to calibrate a user's knowledge before a search session, and assess it after the session, we were able to quantify knowledge gain. The collected data has been released for the purpose of supporting research in the field\footnote{http://l3s.de/~yu/knowledge\_in\_search/}.

Our investigation revealed a significant effect of information need on user queries and navigational patterns, but no direct effect on the knowledge gain. Users on average exhibited a higher knowledge gain through search sessions pertaining to topics they were less familiar with. For more details and findings please refer to the original paper~\cite{gadiraju2018chiir}.


\subsection{Towards Predicting User Knowledge Gain in Informational Search Sessions}
\label{sec:predict}

Based on the advanced understanding of the relation between user knowledge and their search behavior, we investigate the possibility of using search activity related features to predict knowledge gain and the knowledge state of a user -- avoiding the need for explicit post-search knowledge assessments~\cite{yu2018predicting}. 

In this work~\cite{yu2018predicting}, we aim at classifying the knowledge state (gain) of a user at the end of a given search session. For the sake of this work, a user's knowledge state with respect to a particular information need is defined as the predicted user capability (accuracy) to correctly respond to a set of test questions about the respective information need. We classify the user knowledge state into 3 classes according to the user capability: low knowledge state, moderate knowledge state and high knowledge state. We hence define the user's knowledge gain as the amount of knowledge state change, and consequently classify the knowledge gain into 3 classes: low knowledge gain, moderate knowledge gain and high knowledge gain.

\textbf{Approach. }We approach the problem with supervised models for classification. To this end, each session is represented by a feature vector, consisting out of 79 features related to: i) query (e.g. number of query terms, query complexity), ii) SERP (e.g. number of clicks, click-through ratio), iii) browsing behaviour (e.g. number of pages viewed, average time stay per page), and iv) mouse movement (e.g. total scroll distance, number of mouseovers). We applied several feature selection techniques on the considered set of feature, and a range of standard models for the classification tasks, namely, Naive Bayes (NB), Logistic Regression (LR), Support Vector Machine (SVM), Random Forrest (RF) and Multilayer Perceptron (MP). 

\textbf{Preliminary Result. }Using the search activity log and the knowledge test data we collected through crowdsourcing (see Section \ref{sec:analyse}), we trained and evaluated our classification models. The experimental results underline that a user's knowledge gain and knowledge state can be modeled based on a user's online interactions observable throughout the search process. 
Through feature analysis, we provide evidence for an improved understanding between individual user behavior and the corresponding knowledge state and change.

\subsection{Future Work}
As part of future work, we aim to reproduce and refine the findings in more varied search sessions, where durations and learning intents are more diverse, also involving considerably longer/shorter search sessions and, for instance, procedural knowledge rather than intents focused on declarative knowledge only. This would provide the opportunity to observe evolution-oriented features, for instance, considering the evolution of queries, their length and complexity. 
In addition, in crowd-based quasi experiments understanding of the actual users is very limited and data collected as such is expected to exhibit a certain amount of noise. For these reasons, we aim at conducting equivalent experiments in more controlled lab environments, where reliable information about both user interactions as well as the actual users can be obtained. 
Furthermore, whereas our previous work has focused on user interaction features, ongoing research investigates resource-centric features which take into account the characteristics of resources involved within the user interactions. 

Potential applications for this work include the consideration of user knowledge and its expected learning progress as part of Web search engines and information retrieval approaches, or within informal learning-oriented search settings, such as libraries or knowledge- and resource-centric online platforms.

\section{Supporting Learning throughout Web Search} \label{sec:support}
The application-oriented objective are concerned with eventually supporting users in their learning tasks through i) optimizing user interaction and interfaces, and ii) enhancing the retrieval and ranking process. In this section, we review the status of existing techniques, and discuss the potential future directions.

\subsection{Related Works}

\textbf{User interface and interaction. }
Learning oriented online platforms (e.g. coursera\footnote{https://www.coursera.org/}, mooc\footnote{http://mooc.org/}, Didactalia\footnote{https://didactalia.net}) have been constantly optimized to improve the learning performance of users. Examples are, for instance, the use of learning dashboards to inform users about his/her learning progress or provide discussion forums to enable collaboration among users.
However, within general-purpose search engines, there is a lack of attention for the support of learning, also due to the general-purpose nature of such environments and the variety of tasks conducted there. A central question for research in that area is whether and how interfaces can be adopted to improve learning performance even in such general-purpose environments. 
An attempt has been made by Arora et al.~\cite{arora2015promoting}, by aiming at improving user engagement in learning oriented search tasks through providing richer representation of retrieved Web documents. Specifically, they explored methods of finding useful semantic concepts within retrieved documents, with the objective of creating improved document surrogates for presentation in the SERP.

\textbf{Retrieval and ranking. }
As current search engines are optimized by considering an information need disregarding the learning intent behind a query, relatively little research has been carried out on optimising retrieval and ranking algorithms towards particular learning needs. For instance, Dave et al. \cite{dave2014computational} discussed the potential of two ranking models with varied objectives (i.e. paragraph retrieval model,  dependency based re-ranking) on enhancing the performance of learning-centric search engines. Recently, Syed and Collins-Thompson~\cite{syed2017retrieval} proposed to optimize the learning outcome of the vocabulary learning task by selecting a set of documents while considering keyword density and domain knowledge of the learner. Their theoretical framework provides a sound basis for furthering the study on learning-oriented retrieval techniques.

\subsection{Future Work}
\textbf{User interface and interaction. }Within general-purpose search engines, there is a lack of attention for the support of learning, also due to the general-purpose nature of such environments and the variety of tasks conducted there. A central question for research in that area is whether and how interfaces can be adopted to improve learning performance even in such general-purpose environments. 
Studies reveal that people engage more in many search tasks involving collaboration with others rather than while searching by themselves~\cite{morris2007collaborating}. To further this investigation and develop tools to support learning by enabling collaboration between users, our ongoing work is concerned with developing a search interface that encourages experienced learners to guide learners who will use the system in the future and assess its impact on the knowledge gain throughout a search session. Suggestions are ranked according to the feedbacks from experienced learners. Throughout large-scale quasi-experiments and facilitated by pre- and post-tests, we aim to quantify the influence of the collaborative search interface on the learning outcome. Future work is concerned with alternative means to improve interfaces and interactions towards increasing the learning outcomes during Web search. 

\textbf{Retrieval and ranking.} In Section \ref{sec:predict}, we have discussed means to infer a user's knowledge state (gain) in online search sessions. On this basis, future work is aimed at optimizing ranking algorithms to recommend resources that fit not only the traditional notion of an information need, but also a user's knowledge state. Whereas traditional ranking algorithms tend to suggest Web documents disregarding, for instance, a user's reading level, improved ranking techniques will favor resources which are neither too easy nor too hard for a particular user's need. This builds on the assumption that, based on the assessment of the relation between a Web resource and user's knowledge gain, a ranking algorithm can recommend resources not only fitting into the user's knowledge state, but also maximizing the user's knowledge gain and learning efficiency.

\section{Conclusions}
This paper has provided an overview of challenges and research approaches towards detecting, understanding and supporting learning throughout Web search. One crucial challenge in this context is the lack of explicit information about users, their learning intent, task or progress throughout an online search session, requiring the utilisation of a wide variety of informal features to derive such information. In particular, supervised machine learning techniques and extensive feature analysis have been deployed as part of previous work, yet works are usually focused on specific learning scenarios, isolated feature sets or single-query scenarios, rather than entire search sessions. 

Another major obstacle is the lack of large-scale datasets to facilitate SAL research by providing both diverse features of user interactions and behavior as well as high-quality ground truth data about the involved users, their knowledge state and knowledge gain throughout the captured search sessions. 

In addition to summarising research challenges and related works, we have introduced some of our own contributions to the respective tasks. These consist of supervised approaches for detecting learning-related (informational) search sessions, for predicting the knowledge state and gain of online users and the preliminary analysis of experimentally obtained search sessions and the correlation of observed variables with user knowledge. Ongoing and future work will expand on these works, consider more varied feature sets, in particular resource-centric features, and will in particular be concerned with obtaining, providing and analysing search session data collected in more controlled lab environments. 
\balance
\bibliographystyle{ACM-Reference-Format}
\bibliography{reference} 

\end{document}